\documentclass[a4paper,fleqn,usenatbib]{mnras}
\usepackage[T1]{fontenc}
\usepackage{ae,aecompl}
\usepackage{graphicx}	
\usepackage{amsmath}	
\usepackage{amssymb}	
\usepackage{multirow}
\usepackage{subfigure}
\usepackage{times}

\title[CMB Self-Calibration with Polarized Foregrounds]{Foreground-Induced Biases in CMB Polarimeter Self-Calibration}

\author[M. H. Abitbol, J. C. Hill, and B. R. Johnson]{Maximilian H. Abitbol$^{1}$\thanks{E-mail:~\url{mha2125@columbia.edu}}, J. Colin Hill$^{2}$, and Bradley R. Johnson$^{1}$\\
$^{1}$Department of Physics, Columbia University, New York, NY, 10027, USA\\
$^{2}$Department of Astronomy, Columbia University, New York, NY, 10027, USA
}

\date{Accepted 2016 January 2. Received 2015 December 18}
\pubyear{2016}

\begin{document}
\label{firstpage}
\pagerange{\pageref{firstpage}--\pageref{lastpage}}
\maketitle

\begin{abstract}
Precise polarisation measurements of the cosmic microwave background (CMB) require accurate knowledge of the instrument orientation relative to the sky frame used to define the cosmological Stokes parameters. Suitable celestial calibration sources that could be used to measure the polarimeter orientation angle are limited, so current experiments commonly `self-calibrate.' The self-calibration method exploits the theoretical fact that the $EB$ and $TB$ cross-spectra of the CMB vanish in the standard cosmological model, so any detected $EB$ and $TB$ signals must be due to systematic errors. However, this assumption neglects the fact that polarized Galactic foregrounds in a given portion of the sky may have non-zero $EB$ and $TB$ cross-spectra. If these foreground signals remain in the observations, then they will bias the self-calibrated telescope polarisation angle and produce a spurious $B$-mode signal. In this paper we estimate the foreground-induced bias for various instrument configurations and then expand the self-calibration formalism to account for polarized foreground signals. Assuming the $EB$ correlation signal for dust is in the range constrained by angular power spectrum measurements from Planck at 353~GHz (scaled down to 150~GHz), then the bias is negligible for high angular resolution experiments, which have access to CMB-dominated high $\ell$ modes with which to self-calibrate. Low-resolution experiments observing particularly dusty sky patches can have a bias as large as $0.5^\circ$. A miscalibration of this magnitude generates a spurious $BB$ signal corresponding to a tensor-to-scalar ratio of approximately $r\sim2\times10^{-3}$, within the targeted range of planned experiments. 
\end{abstract}


\section{Introduction}
The cosmic microwave background (CMB) is a primordial bath of photons that permeates all of space and carries an image of the Universe as it was 380,000 years after the Big Bang. Physical processes that operated in the early Universe left various imprints in the CMB. These imprints appear today as angular anisotropies, and the primordial angular anisotropies have proven to be a trove of cosmological information. The precise characterization of the intensity (or temperature) anisotropy of the CMB has helped reveal that space-time is flat, the Universe is 13.8 billion years old, and the energy content of the Universe is dominated by cold dark matter and dark energy~\citep{bennett_2013,planck2015params}. The associated `$E$-mode' polarisation anisotropy signal has been observed at the theoretically expected level~\citep{bennett_2013,planck2015power,quiet_2013,naess_2014,crites_2015}. Experimental CMB polarisation research is currently focused on (i) searching for the primordial `$B$-mode' polarisation anisotropy signal from inflationary gravitational waves (IGW)~\citep{Zaldarriaga,kamionkowski_1997} and (ii) characterizing the detected non-primordial $B$-mode signal generated when $E$-modes are gravitationally lensed by large-scale structures in the Universe~\citep{sptpol2013,actpol2014,bicep2_2014a,bicep2keck95,bicep2keck150,sptpolb2015,abazajian_2015a,bicep2keckplanck}. A key challenge for $B$-mode studies is disentangling foreground signals from CMB observations because they can appreciably bias the results in a variety of ways~\citep{chill2014}. In this paper we address biases to polarimeter calibration. 

\begin{figure*}
\subfigure{\includegraphics[scale=0.43]{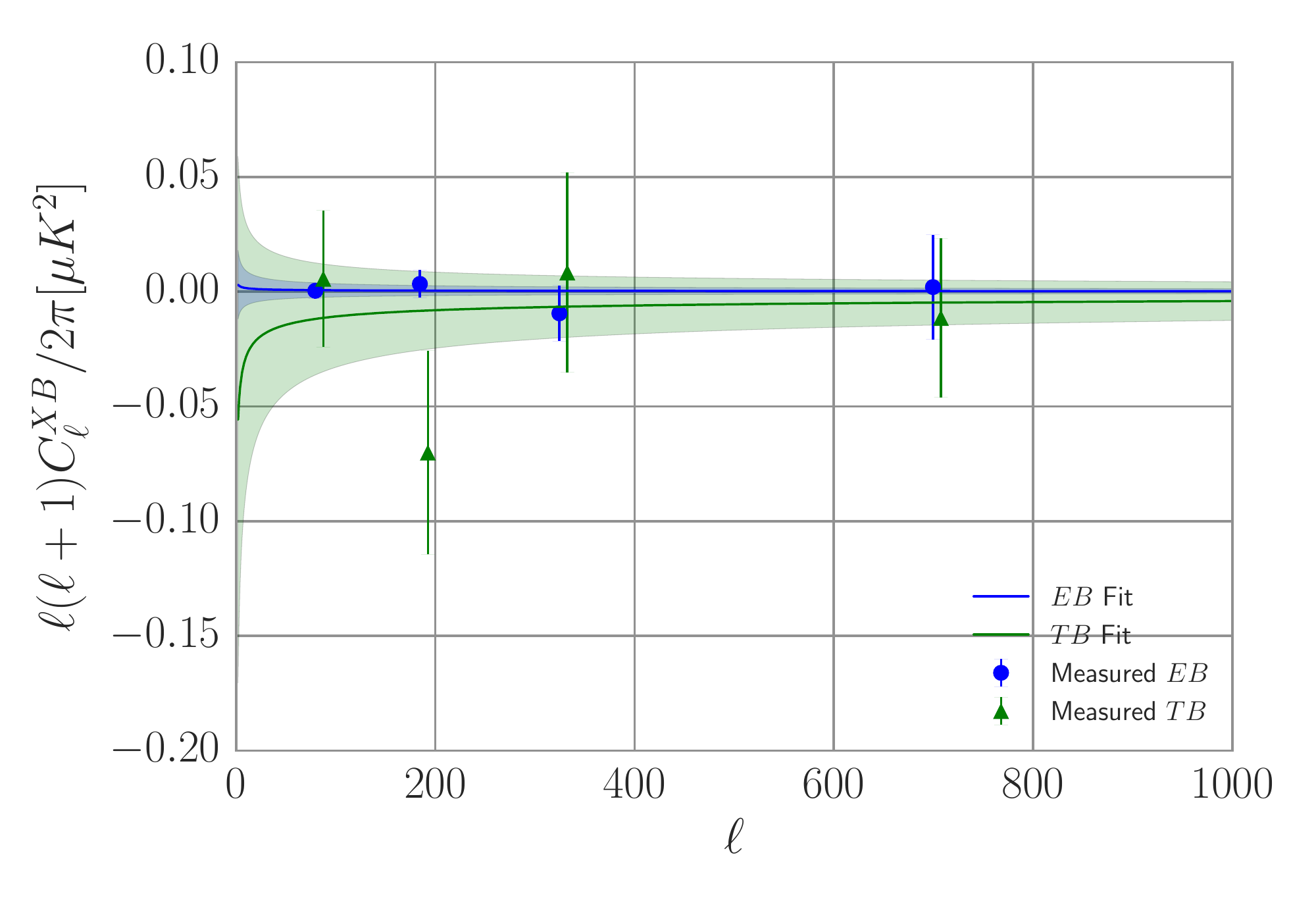} \label{fig:ebtbspec}}\hfill
\subfigure{\includegraphics[scale=0.43]{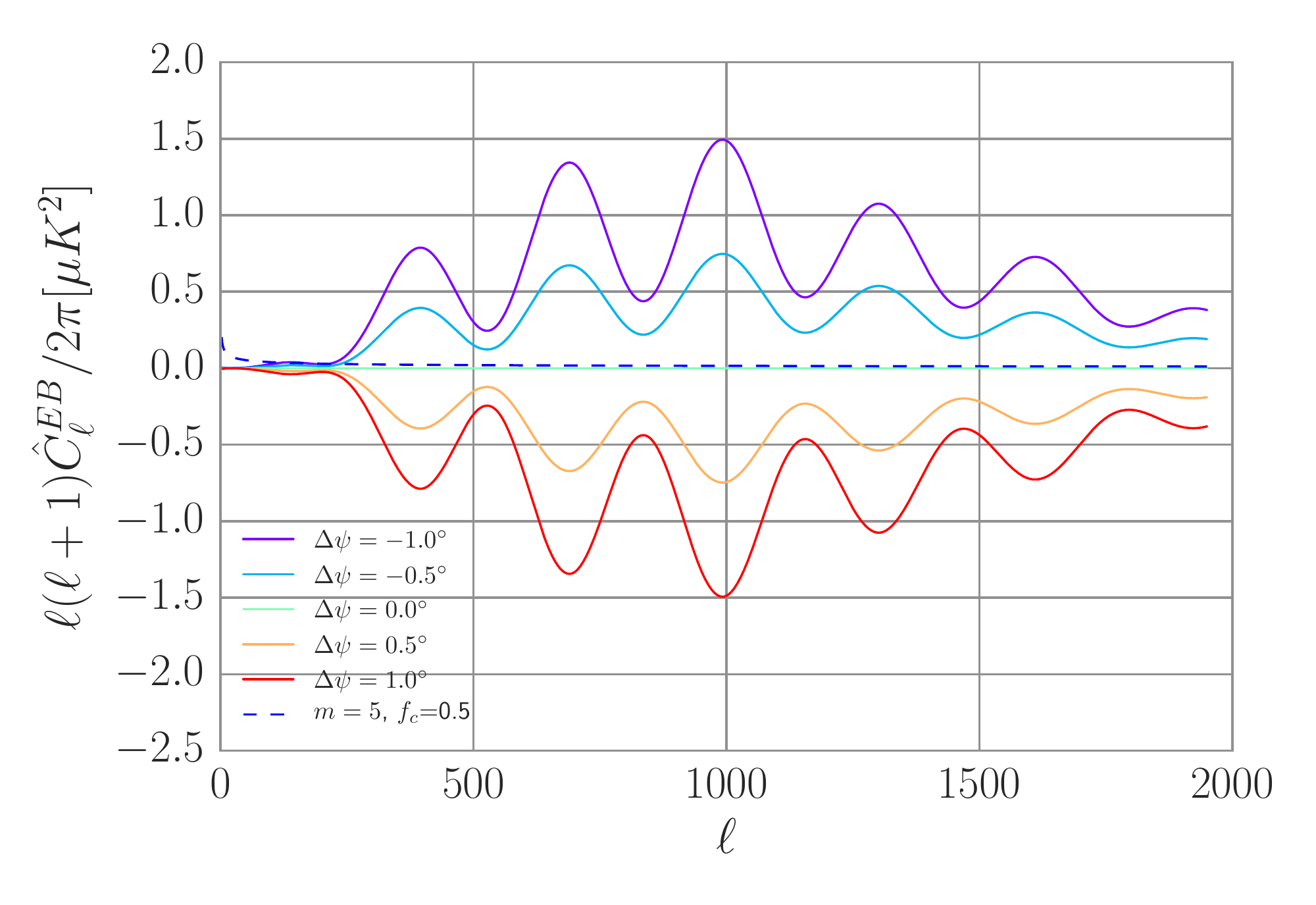} \label{fig:obseb}}
\caption{(a: Left) Measured $EB$ and $TB$ dust cross-spectra with best-fitting power laws. The blue and green data points show the $EB$ and $TB$ dust cross-spectra, respectively, as measured on Planck 353~GHz data in the BICEP2 region using PolSpice and scaled to 150~GHz using the dust grey-body spectrum (plotted with a slight offset for clarity). The solid lines are the best-fitting power laws with a fixed index and the shading represents the uncertainty of the best-fitting amplitude. (b: Right) Rotated CMB $EB$ cross-spectra and dust $EB$ cross-spectra as a correlation fraction. The solid rainbow lines show the rotated $EB$ spectra given an experiment observing only the CMB but misaligned by various angles $\Delta\psi$. The dashed line is an upper bound on the dust $EB$ spectrum, determined by a correlated fraction of the $E$- and $B$-mode power (see Equation~\ref{eq:correl} with $m=5$ and $f_c=0.5$). At $\ell\gtrsim300$ even the brightest dust $EB$ spectrum is negligible compared to the rotated CMB spectra for $\Delta\psi>0.5^{\circ}$.}
\end{figure*}
Precise measurements of the polarisation properties of the CMB require accurate knowledge of the relationship between the instrument frame and the reference frame on the sky that is used to define the cosmological Stokes parameters. We refer to this relative orientation angle as the polarisation angle of the telescope, $\psi=\psi_{\rm design}+\Delta\psi$, where $\psi_{\rm design}$ is the intended orientation and $\Delta\psi$ is a small misalignment. Calculations show that $\psi$ must be measured to arcminute precision for IGW searches targeting tensor-to-scalar ratios $r\lesssim0.01$~\citep{hu2003,odea,miller2009}. Ideally, celestial sources would be used to measure the polarisation angle~\citep{aumont2010,matsumura2010,naess_2014,polarbear_2014}. At millimetre wavelengths, the best celestial source appears to be Tau A, though it is not ideal because (i) it is not bright enough to give a high signal-to-noise ratio measurement with a short integration time, (ii) the source is extended with a complicated polarisation intensity morphology, (iii) the millimetre-wave spectrum of Tau A is not precisely known, which is important for polarimeters that have frequency-dependent performance, and (iv) Tau A is not observable from Antarctica where many ground-based and balloon-borne experiments are sited (see~\cite{johnson_2015} and references therein). Since an ideal celestial calibration source does not exist, many current experiments~\citep{bicep1_calib,bicep1result,naess_2014,polarbear_2014,bicep2_instrument} use a `self-calibration' method~\citep{keatingself}. 

The self-calibration method exploits the fact that the $EB$ and $TB$ cross-spectra of the CMB vanish for parity-conserving inflaton fields~\citep{Zaldarriaga}, so any detected $EB$ and $TB$ signals are interpreted as systematic effects that can be used to de-rotate instrument-induced biases. However, this assumption neglects the fact that polarized Galactic foregrounds in a given portion of the sky may have non-zero $EB$ and $TB$ cross-spectra. If these foreground signals remain in the observations, then they will bias the derived polarisation angle and produce a spurious $BB$ signal even if the instrument was perfectly mechanically aligned before self-calibration~\citep{hu2003,shimon2008,yadav2010}. 

In this paper, we consider the case where the $EB$ and $TB$ cross-spectra for Galactic dust are non-zero, and estimate the foreground-induced bias produced by different levels of polarized dust intensity for various instrument configurations. We then expand the formalism to mitigate the effect of polarized foreground signals. The paper is structured as follows. In Section~\ref{secdata}, we discuss the data used to establish the likely range of polarized foreground signals. In Section~\ref{seccalib}, we review the self-calibration method and add foregrounds to the rotated power spectra. Miscalibration results for different levels of dust and instrument designs are presented in Section~\ref{secresult}. We then estimate the bias on $r$ due to miscalibration in Section~\ref{secspur}.  Section~\ref{secfore} corrects the self-calibration formalism to account for foregrounds in the observations. We show that this recovers the telescope polarisation angle at the cost of increased statistical error on $\Delta\psi$. We summarize and conclude in Section~\ref{secdisc}.

\section{Methods}
\subsection{Estimating Foreground Power Spectra}\label{secdata}
To perform the self-calibration calculation and estimate the foreground bias we require CMB and foreground power spectrum measurements. We use CAMB\footnote{\url{http://camb.info}} to generate theoretical CMB power spectra~\citep{Lewis:1999bs}, with the Planck best-fitting $\Lambda$CDM cosmological parameters~\citep{planck2015params}. We include gravitational lensing and set the tensor-to-scalar ratio $r=0.0$.

To assess the impact of dust foregrounds on the self-calibration procedure, we require estimates of realistic dust $EE$, $TE$, $TB$, $EB$, and $BB$ power spectra. For this purpose, we consider the BICEP2 field, in which foregrounds have been particularly well-studied (e.g., ~\cite{fuskeland2014,chill2014,planck2013foregrounds,bicep2keckplanck,bicep2keck95}), and in which the foreground levels are low but non-negligible. Later we also consider different scenarios for the dust amplitude. To measure the power spectra, we use the Planck 353 GHz $T$, $Q$, and $U$ half-mission split maps and an angular mask approximating the BICEP2 region from~\cite{chill2014}. We also apply a polarized point source mask constructed from Planck High Frequency Instrument data. The combined mask is apodized using a Gaussian with FWHM = 30 arcmin, yielding an effective sky fraction $f_{\rm sky} = 0.013$.  Our power spectrum estimator is based on PolSpice~\citep{polspice2004}, with parameters calibrated using 100 simulations of polarized dust power spectra consistent with recent Planck measurements~\citep{planck2013foregrounds}. The power spectra are estimated from cross-correlations of the Planck half-mission splits, such that no noise bias is present in the results. We bin the measured power spectra in four multipole bins matching those used in~\cite{planck2013foregrounds}, spanning $40 < \ell < 1000$. Error bars are estimated in the Gaussian approximation from the auto-power spectra of the half-mission splits. We then re-scale the 353 GHz measurements to 150 GHz using the best-fitting greybody dust SED from~\cite{planck2013foregrounds}, which corresponds to a factor of $0.041$ for polarisation and $0.043$ for temperature.

Our measured $BB$ power spectrum is consistent with that measured in the BICEP2 patch in~\cite{planck2013foregrounds} (small deviations are expected due to the slightly different masks employed). The measured $EB$ and $TB$ power spectra are shown in Fig.~\ref{fig:ebtbspec}. Both spectra are consistent with zero. The observed amplitudes are smaller than those measured in~\cite{planck2013foregrounds} for $EB$ and $TB$ spectra on large sky fractions containing more dust, as expected (see Appendix~\ref{sec:A} here and figs. B.2 and B.3 of~\cite{planck2013foregrounds}).  We note that even on the large sky fractions studied in~\cite{planck2013foregrounds}, the dust $EB$ and $TB$ spectra are generally consistent with zero, except for masks with $f_{\rm sky} \gtrsim 0.5$, which show evidence of a signal on roughly degree angular scales. We take the results measured in the BICEP2 patch as fiducial dust power spectra and consider variations around this scenario below. For simplicity, we fit a simple power-law template to all dust power spectra (see below), such that each spectrum is completely characterized by an overall amplitude. We then vary the amplitudes to produce two additional sets of dust power spectra to represent different possible observations:
\begin{equation}
C^{dust,XY}_{\ell} = A^{XY}\bigg(\frac{\ell}{80}\bigg)^{-2.42}
\label{eq:dustamp}
\end{equation}
\begin{equation}
C^{dust,XY}_{\ell,mult} = m C^{dust,XY}_{\ell}
\label{eq:dlevel}
\end{equation}
\begin{equation}
C^{dust,ZB}_{\ell,corr} = f_c\sqrt{C^{dust,ZZ}_{\ell,mult} C^{dust,BB}_{\ell,mult}} \,,
\label{eq:correl}
\end{equation}
\noindent where $A^{XY}$ is the best-fitting amplitude, $m$ is a multiplicative factor, $f_c$ is a correlation fraction, $X,Y\in\{T,E,B\}$ and $Z\in\{T,E\}$.

Data set $1$ is calculated by fitting for the amplitude of a power law spectrum to each of the dust power spectra, with a fixed index $\beta=-2.42$, as given by Equation~\ref{eq:dustamp}. This is motivated by the Planck foreground analysis which finds the dust power spectra to be consistent with a power law in $\ell$~\citep{planck2015foregrounds}.

In data set $2$, we increase the amplitude of all dust power spectra by an overall multiplicative factor, $m$, given by Equation~\ref{eq:dlevel}. This represents measurements on patches larger and dustier than the BICEP2 region. 

In data set $3$, we write the $EB$ and $TB$ dust spectra as a correlated fraction, $f_c$, of the $EE$ and $BB$ and $TT$ and $BB$ spectra, respectively, as given by Equation~\ref{eq:correl}. We use the correlation fraction to explore the possibility of proportionally large $EB$ and $TB$ cross-spectra while imposing the constraint that they do not exceed the level of the $EE$ and $BB$ or $TT$ and $BB$ power, respectively. The dashed lines in Fig.~\ref{fig:obseb} show the data set $3$ $EB$ dust cross-spectrum. These data sets provide realistic upper bounds on the observed dust power spectra at 150~GHz. We list the amplitude of the dust cross-spectra in each case in Table~\ref{tab:dustamps}. 

\subsection{Review of Self-Calibration Procedure}
\label{seccalib}
Following the self-calibration procedure of~\cite{keatingself}, a miscalibration of the instrument polarisation angle, $\psi_{design}$, by an amount $\Delta\psi$ results in a rotation of the observed Stokes vector and thus the observed Stokes parameters, $\hat{Q}(\textit{\textbf{p}})$ and $\hat{U}(\textit{\textbf{p}})$, as given by Equations~\ref{eq:qu} and~\ref{eq:qu2}:
\begin{equation}
\hat{Q}(\textit{\textbf{p}}) = \cos(2\Delta\psi) Q(\textit{\textbf{p}}) - \sin(2\Delta\psi) U(\textit{\textbf{p}})
\label{eq:qu}
\end{equation}
\begin{equation}
\hat{U}(\textit{\textbf{p}}) = \sin(2\Delta\psi) Q(\textit{\textbf{p}}) + \cos(2\Delta\psi) U(\textit{\textbf{p}}) \,,
\label{eq:qu2}
\end{equation}
\noindent where $Q(\bf{p})$ and $U(\bf{p})$ are the sky-synchronous linear polarisation Stokes parameters, and $\textit{\textbf{p}}$ denotes the pointing on the sky (note we use the CMB convention for the polarisation angle direction). The observed $\hat{E}(\textit{\textbf{l}})$ and $\hat{B}(\textit{\textbf{l}})$ modes are then rotated from the sky-synchronous $E(\textit{\textbf{l}})$ and $B(\textit{\textbf{l}})$ modes as given by Equations~\ref{eq:ebmodes} and~\ref{eq:ebmodes2}:
\begin{equation}
\hat{E}(\textit{\textbf{l}}) = \cos(2\Delta\psi)E(\textit{\textbf{l}}) + \sin(2\Delta\psi)B(\textit{\textbf{l}})
\label{eq:ebmodes}
\end{equation}
\begin{equation}
\hat{B}(\textit{\textbf{l}}) = -\sin(2\Delta\psi)E(\textit{\textbf{l}}) + \cos(2\Delta\psi)B(\textit{\textbf{l}}) \,,
\label{eq:ebmodes2}
\end{equation}
\noindent where $\textit{\textbf{l}}$ is the conjugate variable to $\textit{\textbf{p}}$. To determine the best-fitting misalignment angle, $\Delta\psi$, we minimize the variance between the rotated spectra and theoretical CMB spectra and thus maximize the likelihood functions given by Equations~\ref{eq:likelihoods} and~\ref{eq:likelihoods2}, which are analytically solvable. We define $f_{\rm sky}$ as the observed sky fraction, $\Delta_X$ as the observation noise, $\Theta_{\rm FWHM}$ as the telescope beam full-width at half-maximum, and let the subscript and superscript $X\in\{T,E,B\}$.
\begin{equation}
\mathcal{L}_{EB}(\Delta\psi) \propto \exp\Bigg[-\sum_{\ell}\frac{\Big(\hat{C}^{EB}_{\ell} + \frac{1}{2}\sin(4\Delta\psi)\big(C^{EE}_{\ell} - C^{BB}_{\ell}\big) \Big)^2}{2\big(\delta \hat{C}^{EB}_{\ell} \big)^2}\Bigg]
\label{eq:likelihoods}
\end{equation}
\begin{equation}
\mathcal{L}_{TB}(\Delta\psi) \propto \exp\Bigg[-\sum_{\ell}\frac{\Big(\hat{C}^{TB}_{\ell} + \sin(2\Delta\psi)C^{TE}_{\ell}\Big)^2}{2\big(\delta \hat{C}^{TB}_{\ell} \big)^2}\Bigg]
\label{eq:likelihoods2}
\end{equation}
\begin{equation}
\big(\delta\hat{C}^{EB}_{\ell}\big)^2 = \frac{1}{(2\ell+1)f_{\rm sky}}\hat{C}^{EE, tot}_{\ell}\hat{C}^{BB, tot}_{\ell}
\label{eq:delta}
\end{equation}
\begin{equation}
\big(\delta\hat{C}^{TB}_{\ell}\big)^2 = \frac{1}{(2\ell+1)f_{\rm sky}}\hat{C}^{TT, tot}_{\ell}\hat{C}^{BB, tot}_{\ell}
\label{eq:delta2}
\end{equation}
\begin{equation}
\hat{C}^{XX,tot}_{\ell} = \hat{C}^{XX}_{\ell} + \Delta^2_X e^{\ell^2\Theta^2_{\rm FWHM}/(8\ln2)}
\end{equation}
\begin{figure*}
\centering
\subfigure{\includegraphics[scale=0.43]{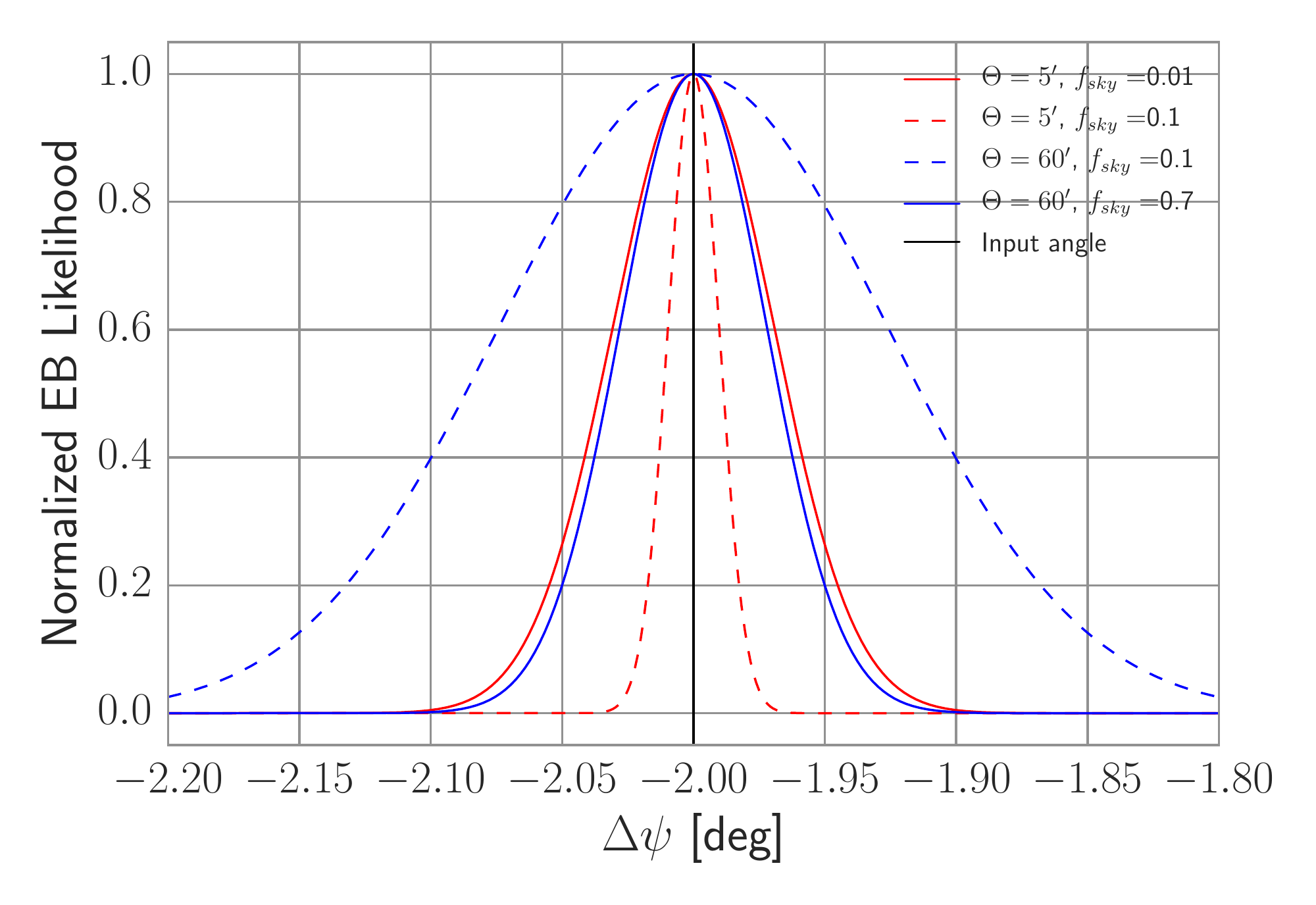}\label{fig:beamfskycalib}}\hfill
\subfigure{\includegraphics[scale=0.43]{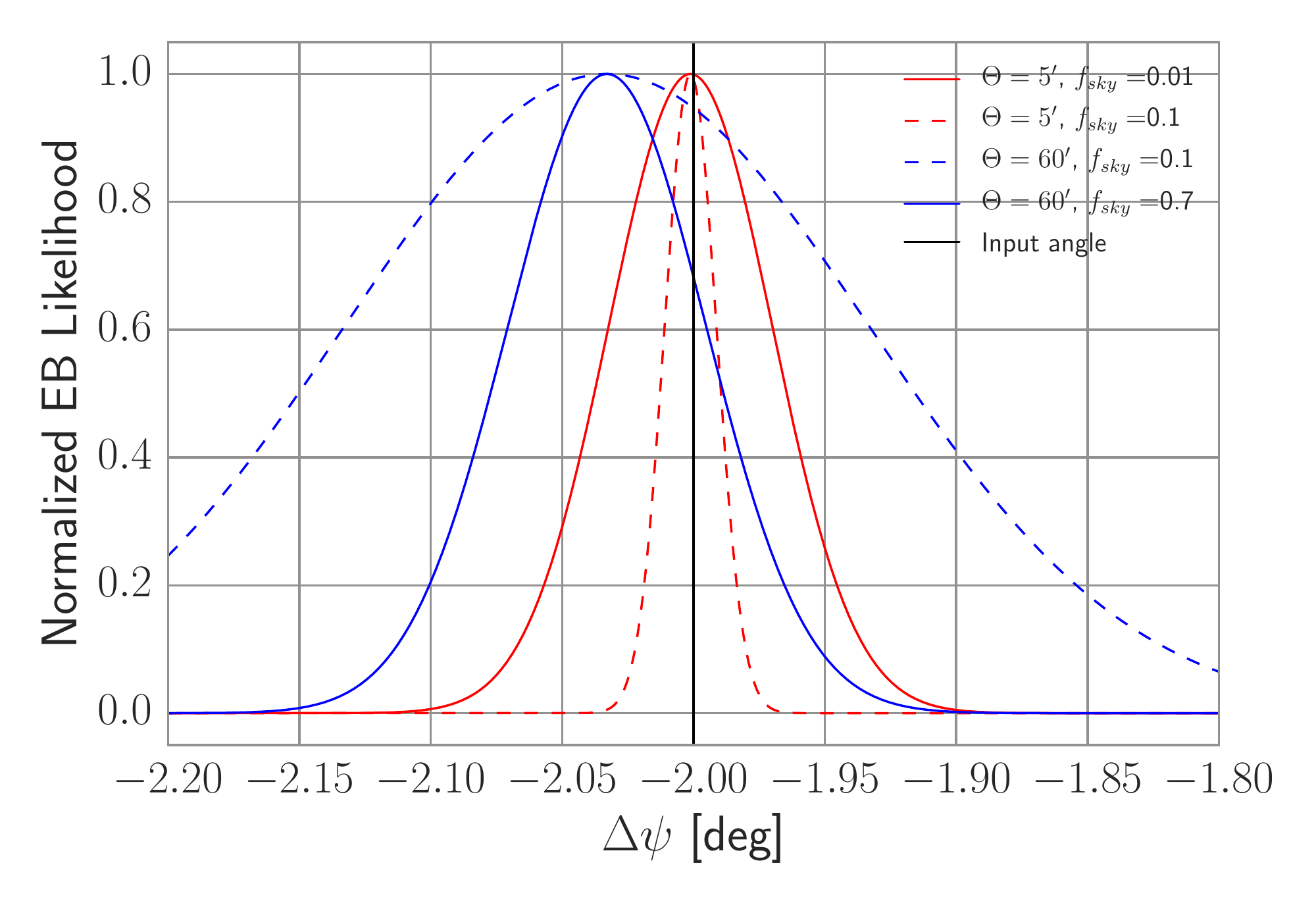} \label{fig:beamfskydustcalib}}
\caption{(a: Left) Clean sky $EB$ likelihoods for various instrument configurations (see Section~\ref{seccmbonly}). Self-calibration likelihood for recovering $\Delta\psi_{EB}$ given an experiment misaligned by $\Delta\psi_{\rm in}=-2^{\circ}$. We exclude dust and consider only the effects of different beam sizes and sky fractions. We test experiment configurations with $\Theta_{\rm FWHM} = 5^{\prime}$ and $60^{\prime}$ and $f_{\rm sky} = 0.01$, $0.1$, and $0.7$. Red curves show the likelihood for $5^{\prime}$ resolution experiments and blue curves show $60^{\prime}$ resolution experiments. The two experiment configurations with nearly the same likelihood will be referred to as experiments $E_{HR}$ and $E_{LR}$ for the high-resolution and low-resolution configurations, respectively. (b: Right) Dusty sky $EB$ likelihoods for various instrument configurations (see Section~\ref{secdust1}). The same experiment specifications as Fig.~\ref{fig:beamfskycalib} but including dust in the rotated power spectrum. Dust dominates the polarized CMB at low multipoles and thus weakens the self-calibration procedure for low-resolution experiments. There is a small bias in the recovered alignment angle for large beam experiments.}
\end{figure*}
\noindent The hat on $\hat{C}_{\ell}$ denotes rotated angular power spectra, which include foregrounds and a rotation angle in the model, and represent, in this paper, the power spectra that would be measured by an experiment. $C_{\ell}$ represents theoretical CMB power spectra. The rotated power spectra are given by Equations~\ref{eq:observed} -~\ref{eq:observed4} with the foregrounds given by $C^{fg}_{\ell}$.
\begin{flalign}
\hat{C}^{EE}_{\ell} = & \sin^2(2\Delta\psi)\big(C^{BB}_{\ell} + C^{fg,BB}_{\ell}\big) + \cos^2(2\Delta\psi) \nonumber \\ &  \times\big(C^{EE}_{\ell} + C^{fg,EE}_{\ell}\big) + \sin(4\Delta\psi)C^{fg,EB}_{\ell}
\label{eq:observed}
\end{flalign}
\begin{flalign}
\hat{C}^{BB}_{\ell} = & \cos^2(2\Delta\psi)\big(C^{BB}_{\ell} + C^{fg,BB}_{\ell}\big) + \sin^2(2\Delta\psi) \nonumber \\ & \times\big(C^{EE}_{\ell} + C^{fg,EE}_{\ell}\big) - \sin(4\Delta\psi)C^{fg,EB}_{\ell}
\label{eq:observed1}
\end{flalign}
\begin{flalign}
\hat{C}^{TE}_{\ell} = & \cos(2\Delta\psi)\big(C^{TE}_{\ell} + C^{fg,TE}_{\ell}\big) + \sin(2\Delta\psi) C^{fg,TB}_{\ell}
\label{eq:observed2}
\end{flalign}
\begin{flalign}
\hat{C}^{TB}_{\ell} = & \cos(2\Delta\psi) C^{fg,TB}_{\ell} -\sin(2\Delta\psi)\big(C^{TE}_{\ell} + C^{fg,TE}_{\ell}\big)
\label{eq:observed3}
\end{flalign}
\begin{flalign}
\hat{C}^{EB}_{\ell} = & \frac{1}{2}\sin(4\Delta\psi)\big(C^{BB}_{\ell} - C^{EE}_{\ell} + C^{fg,BB}_{\ell} - C^{fg,EE}_{\ell}\big) \nonumber \\ 
& + \cos(4\Delta\psi) C^{fg,EB}_{\ell}
\label{eq:observed4}
\end{flalign}
\noindent We use dust power spectra as defined by Equations~\ref{eq:dustamp} -~\ref{eq:correl} as the foreground spectra. Fig.~\ref{fig:obseb} shows the rotated $EB$ spectrum, without dust, for various rotation angles. Fig.~\ref{fig:rotatedpower} shows the CMB, dust, and rotated power spectra as well as noise for a fiducial experiment design. 

Once $\Delta\psi$ is found it can be corrected for by a rotation of $-\Delta\psi$ applied to the measured $Q$ and $U$ maps, however any non-zero $EB$ or $TB$ foreground power will bias the calibration angle, as shown in the next Section. We write a complete formalism that takes the foregrounds into account in the likelihood itself in Section~\ref{secfore}.

\begin{table}
\begin{center}
\begin{tabular}{|c|c|c|c|} 
 \hline
  \multicolumn{4}{|c|}{$\Delta\psi_{EB}$ [degrees]} \\ \hline
  $\Theta_{\rm FWHM}$ & $f_{\rm sky}$ & CMB Only & CMB + Dust \\  \hline
  \multirow{2}{*}{$5.0^{\prime}$} & $0.01$ & $-2.00\pm0.03$ & $-2.00\pm0.03$ \\
  & $0.1$ & $-2.00\pm0.01$ & $-2.00\pm0.01$ \\ \hline
  \multirow{2}{*}{$60.0^{\prime}$} & $0.1$ & $-2.00\pm0.07$ & $-2.03\pm0.10$ \\
  & $0.70$ & $-2.00\pm0.03$ & $-2.03\pm0.04$ \\ \hline
  \multicolumn{4}{|c|}{$\Delta\psi_{TB}$ [degrees]} \\ \hline
  \multirow{2}{*}{$5.0^{\prime}$} & $0.01$ & $-2.00\pm0.09$ & $-2.00\pm0.09$ \\
  & $0.1$ & $-2.00\pm0.03$ & $-2.00\pm0.03$ \\ \hline
  \multirow{2}{*}{$60.0^{\prime}$} & $0.1$ & $-2.00\pm0.12$ & $-2.03\pm0.13$ \\
  & $0.70$ & $-2.00\pm0.05$ & $-2.03\pm0.06$ \\ \hline
\end{tabular}
\caption{Recovered angle with and without dust (see Fig.~\ref{fig:beamfskycalib} and~\ref{fig:beamfskydustcalib}). Simulated misalignment $\Delta\psi_{\rm in}=-2.0^{\circ}$. The recovered $\Delta\psi$ and $1\sigma$ uncertainties for experiment configurations with different beams and sky coverage, with and without dust in the rotated spectra. The $EB$ calibration is more precise than $TB$ in all scenarios. The uncertainties scale inversely with $f_{\rm sky}$.}
\label{tab:beamfsky}
\end{center}
\end{table}

\begin{figure*}
\centering
\subfigure{\includegraphics[scale=0.43]{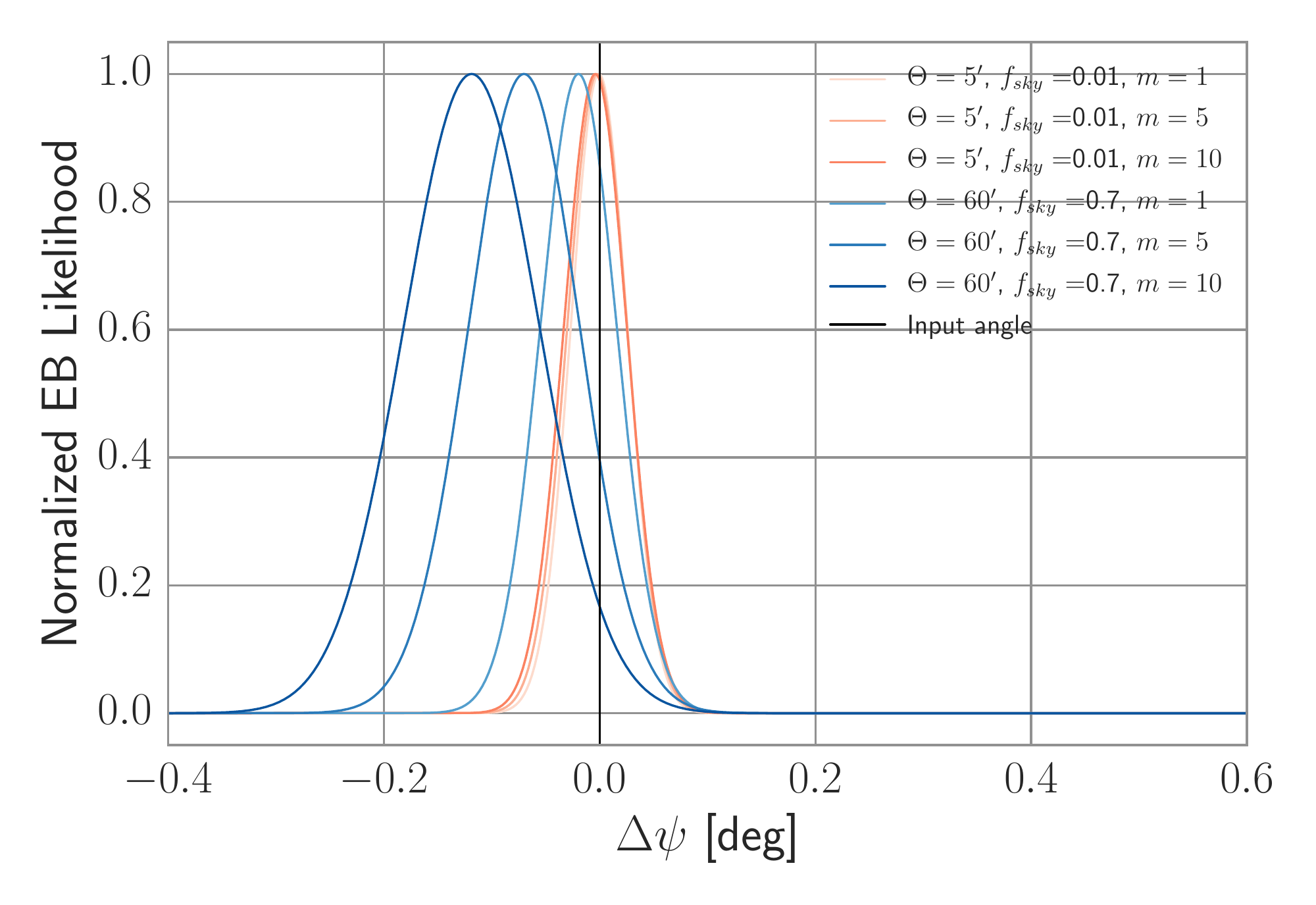} \label{fig:ebdustlevel}}\hfill
\subfigure{\includegraphics[scale=0.43]{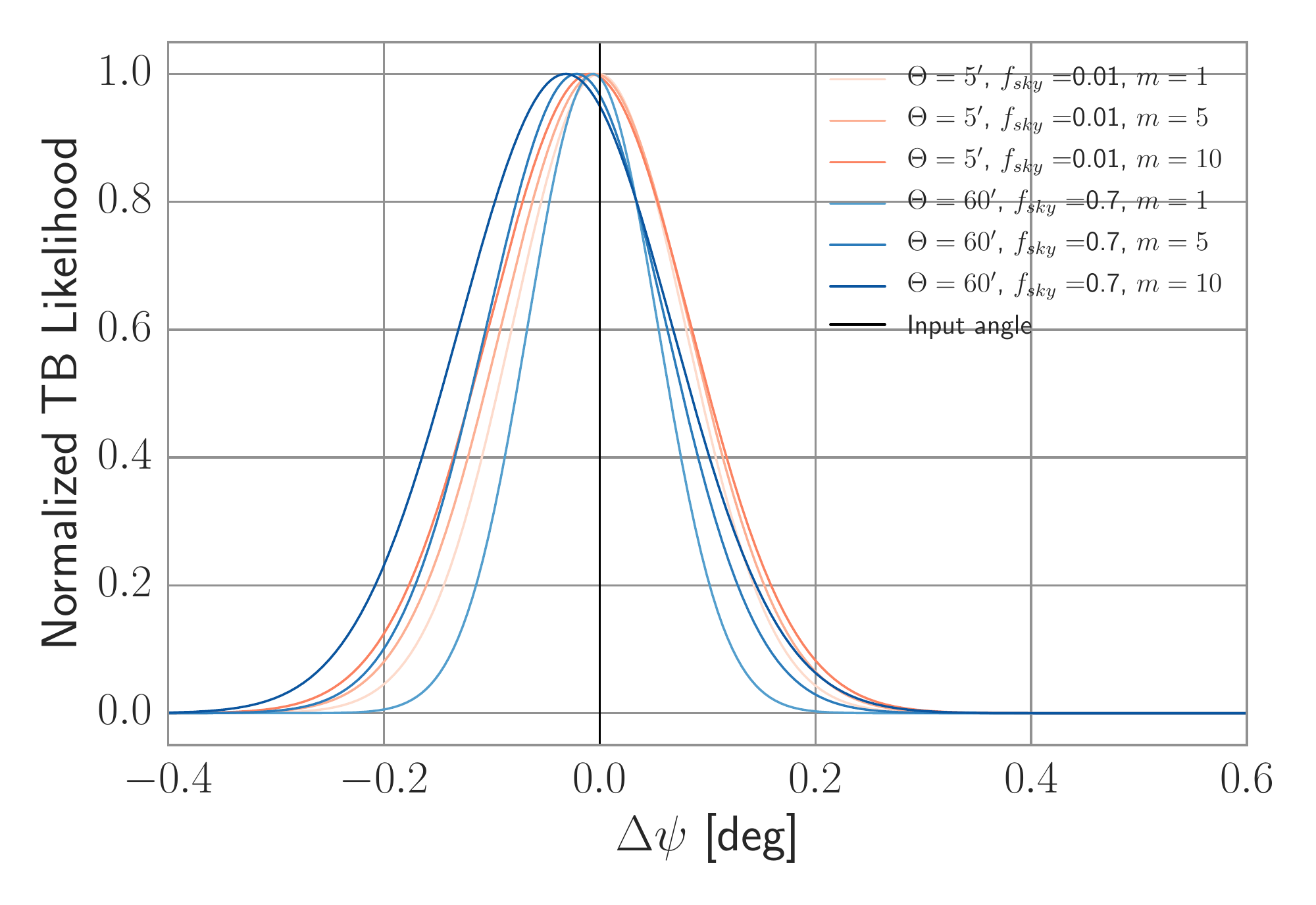}\label{fig:tbdustlevel}}
\caption{(a: Left) $EB$ Likelihood with increased dust level (see Section~\ref{secdust2}). We increase the level of dust power and compare the results for experiment configurations $E_{HR}$ and $E_{LR}$, with the simulated misalignment $\Delta\psi_{in} = 0.0^{\circ}$. The low-resolution experiment is biased and has larger statistical uncertainty than the high-resolution experiment when high levels of foregrounds are observed. The multiplicative factor is $m=1$, $5$, and $10$. (b: Right) $TB$ Likelihood with increased dust level. Same as Fig.~\ref{fig:ebdustlevel} but using $TB$ as a calibrator. Notice the $TB$ calibration has larger uncertainties than $EB$ but is more robust in general to high power foregrounds.}
\end{figure*}

\section{Results}\label{secresult}
\subsection{Foreground Biased Self-Calibration Angle}
We perform the self-calibration procedure to measure the telescope misalignment $\Delta\psi$ using several different instrument configurations and compare the effects of foregrounds in each case. For all experiments we assume an effective instrument noise $\Delta_X=5\mu K$ arcmin. We use three data sets of different dust spectra to characterize the effects of foregrounds on the self-calibration angle. 

For convenience we define $E_{HR}$ as the high-resolution and small sky fraction experiment with $\Theta_{\rm FWHM}=5^{\prime}$ and $f_{\rm sky}=0.01$ and $E_{LR}$ as the low-resolution and large sky fraction experiment with $\Theta_{\rm FWHM}=60^{\prime}$ and $f_{\rm sky}=0.7$. 

\subsubsection{Self-Calibration with CMB Only}\label{seccmbonly}
We reproduce the CMB-only results of~\cite{keatingself} in Fig.~\ref{fig:beamfskycalib} and Table~\ref{tab:beamfsky}, with all the dust spectra set to zero. High angular resolution or large sky fraction experiments have inherently less statistical uncertainty on the self-calibrated angle than low-resolution or small sky fraction experiments. Experiments $E_{HR}$ and $E_{LR}$ have approximately the same constraining power on $\Delta\psi_{EB}$ using the self-calibration procedure on the CMB-only sky.
\subsubsection{Self-Calibration with Dust Measured in BICEP2 Region}\label{secdust1}
We add dust, as measured in the BICEP2 region and fit to a power law, to the rotated spectra as in Equations~\ref{eq:observed} -~\ref{eq:observed4}, and show the results in Fig.~\ref{fig:beamfskydustcalib} and Table~\ref{tab:beamfsky}. The recovered $\Delta\psi$ for experiment $E_{HR}$ is unbiased. However, the dust foreground produces a small bias in the calibration angle of experiment $E_{LR}$ by $\Delta\psi_{\rm in}-\Delta\psi_{\rm out}=0.03^{\circ}$ at $0.75\sigma$ significance. The dust also increases the statistical error of the calibration. A bias of this size is negligible compared to current calibration uncertainties (of order $0.5^{\circ}$), but could prove relevant in the future. Also, the dust power spectra can be larger in other regions of the sky, producing a larger bias, as we show below. 
\subsubsection{Self-Calibration with Brighter Dust Spectra}\label{secdust2}
\begin{table}
\begin{center}
\begin{tabular}{|c|c|c|c|c|} 
 \hline
  \multicolumn{3}{|c|}{Experiment Config.} & \multicolumn{2}{|c|}{$\Delta\psi$ [arcmin]} \\ \hline
  $\Theta_{\rm FWHM}$ & $f_{\rm sky}$ & $m$ & $EB$ & $TB$ \\  \hline
  \multirow{3}{*}{$5.0^{\prime}$} & \multirow{3}{*}{$0.01$} & $1$ & $-0.0\pm1.6$ & $-0.1\pm4.8$ \\
  & & $5$ & $-0.1\pm1.7$ & $-0.2\pm5.2$ \\
  & & $10$ & $-0.2\pm1.7$ & $-0.6\pm5.6$ \\ \hline
  \multirow{3}{*}{$60.0^{\prime}$} & \multirow{3}{*}{$0.70$} & $1$ & $-1.2\pm2.1$ & $-0.4\pm3.6$ \\
  & & $5$ & $-4.2\pm3.0$ & $-1.3\pm5.0$ \\
  & & $10$ & $-7.1\pm3.8$ & $-1.9\pm5.9$ \\
 \hline
\end{tabular}
\caption{Increased dust level by multiplicative factor (see Fig.~\ref{fig:ebdustlevel} and~\ref{fig:tbdustlevel}). Simulated misalignment $\Delta\psi_{\rm in}=0.0^{\circ}$ for experiment configurations $E_{HR}$ and $E_{LR}$ and $m=1$, $5$, and $10$. An experiment observing large portions of the sky near the Galactic plane will observe high levels of dust which can bias the calibration angle, as evident in the second row of the table. Note the units of $\Delta\psi$ are arcminutes.}
\label{tab:levels}
\end{center}
\end{table}
We increase the dust power in all spectra by a multiplicative factor as in Equation~\ref{eq:dlevel}. This is motivated by the fact that we measured the dust spectra on only $1$ per cent of the sky at high Galactic latitude, while larger sky fractions will see more dust. We increase the dust amplitude by up to an order of magnitude, which is consistent with Planck observed dust power on $70$ per cent of the sky.  Fig.~\ref{fig:ebdustlevel} and Table~\ref{tab:levels} illustrate the effect of increasing levels of dust power for experiments $E_{HR}$ and $E_{LR}$. The dust power dominates the CMB at low $\ell$ and thus low resolution experiments using the self-calibration procedure are susceptible to a bias (as large as $1-2\sigma$). The calibration angle for high resolution experiments is robust to strong foregrounds.
\begin{figure*}
\subfigure{\includegraphics[scale=0.43]{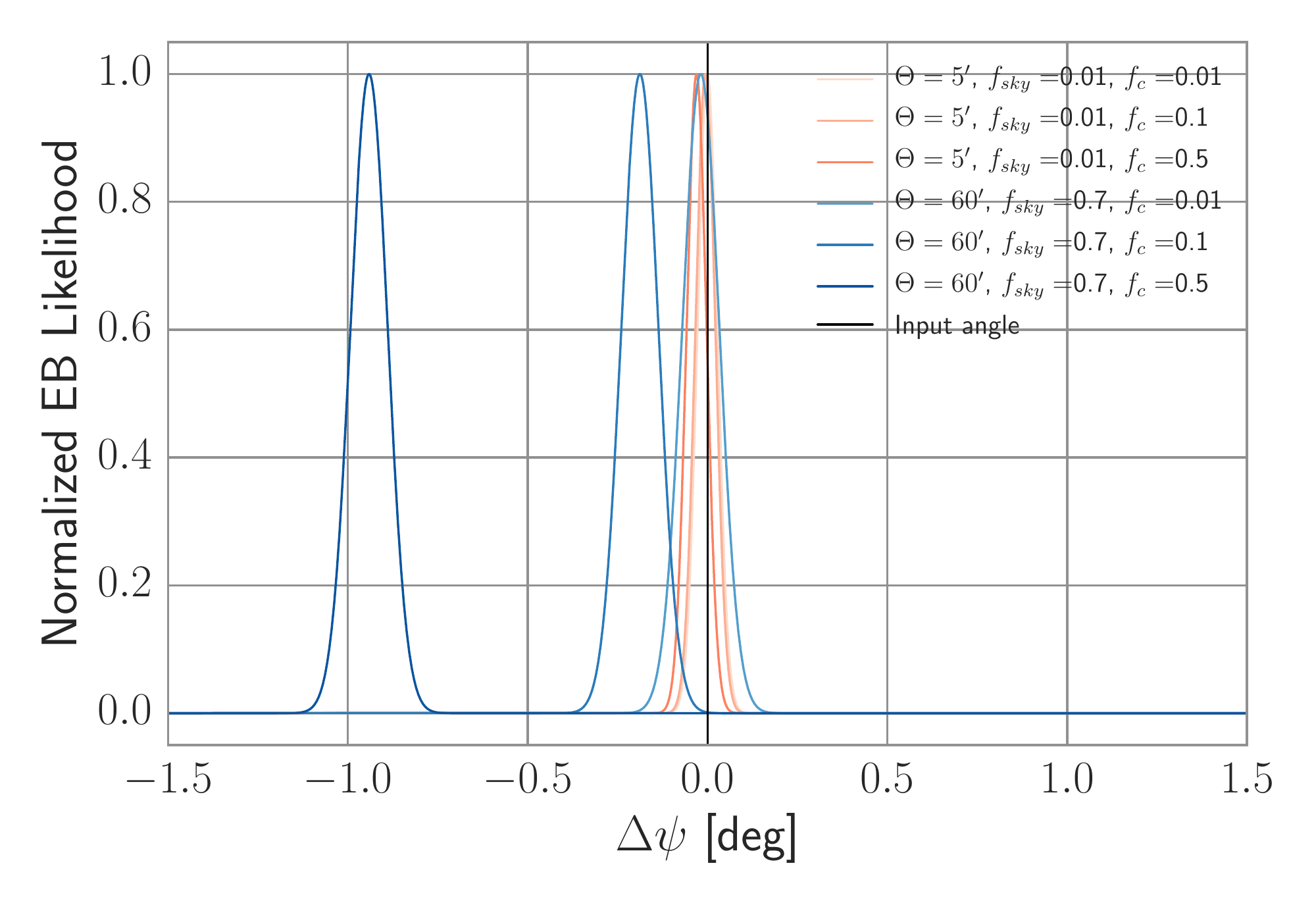}\label{fig:EBfrac}} \hfill
\subfigure{\includegraphics[scale=0.43]{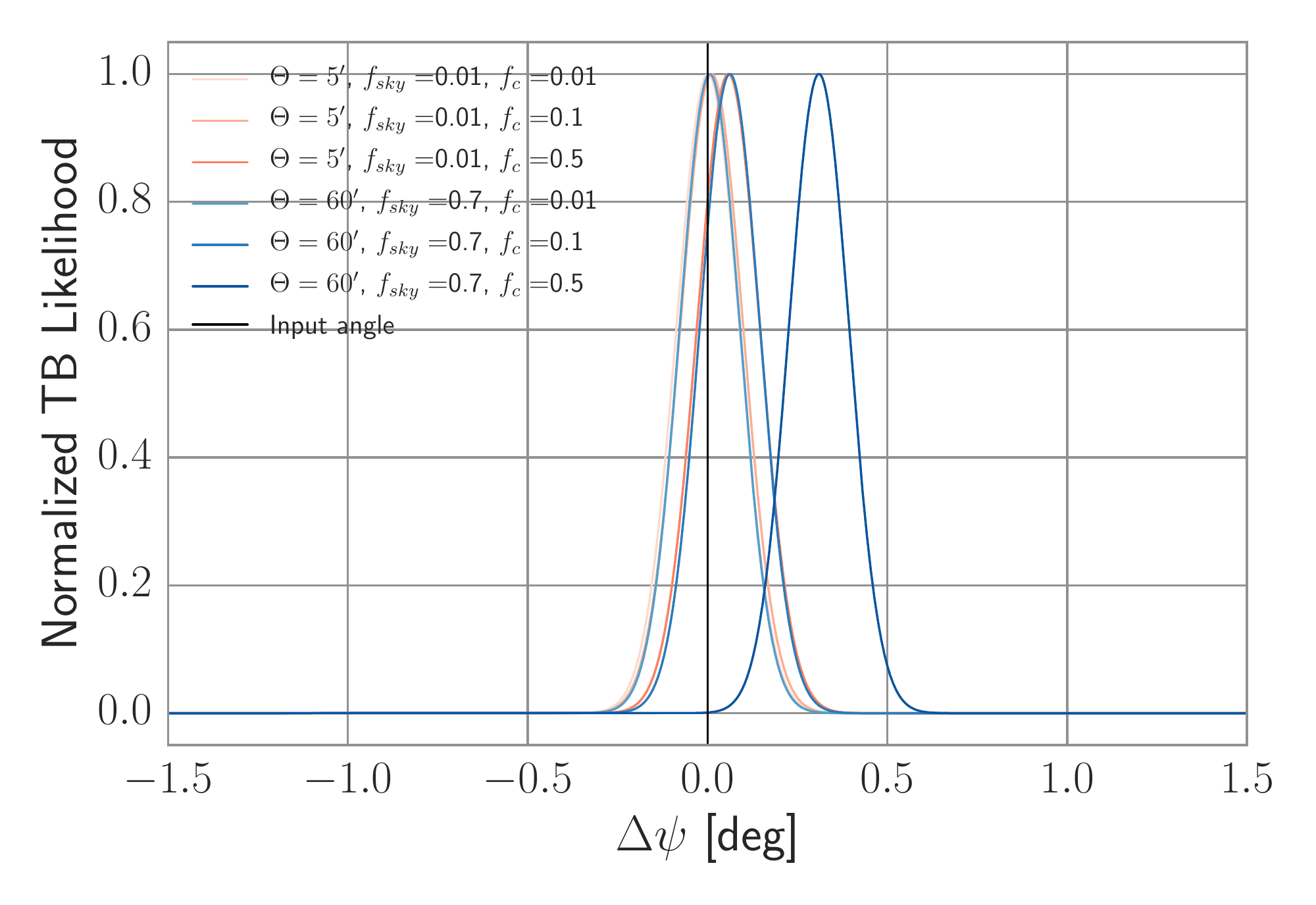}\label{fig:TBfrac}}
\caption{(a: Left) $EB$ likelihood with correlated dust (see Section~\ref{secdust3}). Self-calibration likelihood using correlation fractions to set the $EB$ and $TB$ dust power. We set the overall dust level to be $5\times$ that measured in the BICEP2 region and the correlation fraction $f_c=0.01$, $0.1$ and $0.5$. The low-resolution experiment measures a calibration angle biased by up to $1^{\circ}$. Again the high resolution experiment is robust to foregrounds. (b: Right) $TB$ likelihood with correlated dust. Same as Fig.~\ref{fig:EBfrac} but using $TB$ as a calibrator. The bias in $\Delta\psi_{TB}$ is of the opposite sign of the other $\Delta\psi_{TB}$ results because we have set the $C^{TB}_{\ell}$ dust spectrum to be positive when using the correlation fraction, however it is measured in the BICEP2 region to be slightly negative.}
\end{figure*}

\subsubsection{Self-Calibration with Correlated Dust Spectra}\label{secdust3}
We write the $EB$ and $TB$ spectra as a correlated fraction of the power in $EE$ and $BB$ and $TT$ and $BB$, respectively, as in Equation~\ref{eq:correl}. For simplicity we let the correlation fraction be the same and positive for both $EB$ and $TB$, although the $TB$ spectra measured in the BICEP2 region is slightly negatively correlated. To show an extreme case, we take the dust level to be $5\times$ that measured in the BICEP2 region and then set $EB$ and $TB$ using various correlation fractions, as shown in Fig.~\ref{fig:EBfrac} and~\ref{fig:TBfrac} and Table~\ref{tab:frac}. We plot the $EB$ cross-spectra derived using this method in Fig.~\ref{fig:obseb}.

The self-calibration angle in this scenario can be biased by up to $1^{\circ}$. There are several factors that must conspire together to achieve this bias. First, we used a relatively large beam telescope, although with a large sky fraction. Second, we used dust power $5\times$ that in the BICEP2 region, which is generally only realistic for patches near the Galactic plane. Third, the dust correlation fraction is $50$ per cent which is approximately $100\times$ that measured by Planck. We do not expect this to be observed, although theoretically possible, and thus include it to show an upper bound. Using a small beam eliminates the bias and thus self-calibration for high resolution experiments is robust to bright polarized foregrounds.
\begin{table}
\begin{center}
\begin{tabular}{|c|c|c|c|c|} 
 \hline
  \multicolumn{3}{|c|}{Experiment Config.} & \multicolumn{2}{|c|}{$\Delta\psi$ [arcmin]} \\ \hline
  $\Theta_{\rm FWHM}$ & $f_{\rm sky}$ & $f_c$ & $EB$ & $TB$ \\  \hline
  \multirow{3}{*}{$5.0^{\prime}$} & \multirow{3}{*}{$0.01$} & $0.01$ & $-0.1\pm1.7$ & $0.1\pm5.2$ \\
  & & $0.1$ & $-0.4\pm1.7$ & $0.7\pm5.2$ \\
  & & $0.5$ & $-1.9\pm1.7$ & $3.5\pm5.2$ \\ \hline
  \multirow{3}{*}{$60.0^{\prime}$} & \multirow{3}{*}{$0.70$} & $0.01$ & $-1.1\pm3.1$ & $0.4\pm5.0$ \\
  & & $0.1$ & $-11\pm3.1$ & $3.7\pm5.0$ \\
  & & $0.5$ & $-57\pm3.1$ & $19\pm5.0$ \\ \hline
\end{tabular}
\caption{Increased dust by using a correlation fraction (see Fig.~\ref{fig:EBfrac} and~\ref{fig:TBfrac}). Simulated misalignment $\Delta\psi_{in}=0.0^{\circ}$ for experiment configurations $E_{HR}$ and $E_{LR}$ and setting the dust $EB$ and $TB$ cross-spectra by $m=5$ and $f_c=0.01$, $0.1$, $0.5$. The low-resolution experiment measures significantly biased calibration angles. Disagreement between the $EB$ and $TB$ self-calibration angles would be a sign of foreground biases or other systematic errors.} \label{tab:frac}
\end{center}
\end{table}

\subsection{Self-Calibration Angle Bias and Spurious B-mode Power} 
\label{secspur}
A miscalibration of the telescope angle will generate $B$-mode power from the rotation of $E$-modes into $B$-modes, as shown in Fig.~\ref{fig:EBfrac}. We estimate the tensor-to-scalar ratio from the spurious $B$-mode power for various rotation angles in Table~\ref{tab:rstab}. To estimate the equivalent $r$ we take the rotated $\hat{C}^{BB}_{\ell}$ spectra divided by the $r=1.0$ theoretical $C^{BB}_{\ell}$ spectrum and evaluate at $\ell=80$, for a given angle $\Delta\psi$. We have neglected dust in the calculation as adding dust can produce an additional bias (i.e., we assume high-frequency data is used to clean polarized dust from the maps). We have also excluded lensing $B$-mode power in the calculation.

\begin{figure}
	\includegraphics[width=\columnwidth]{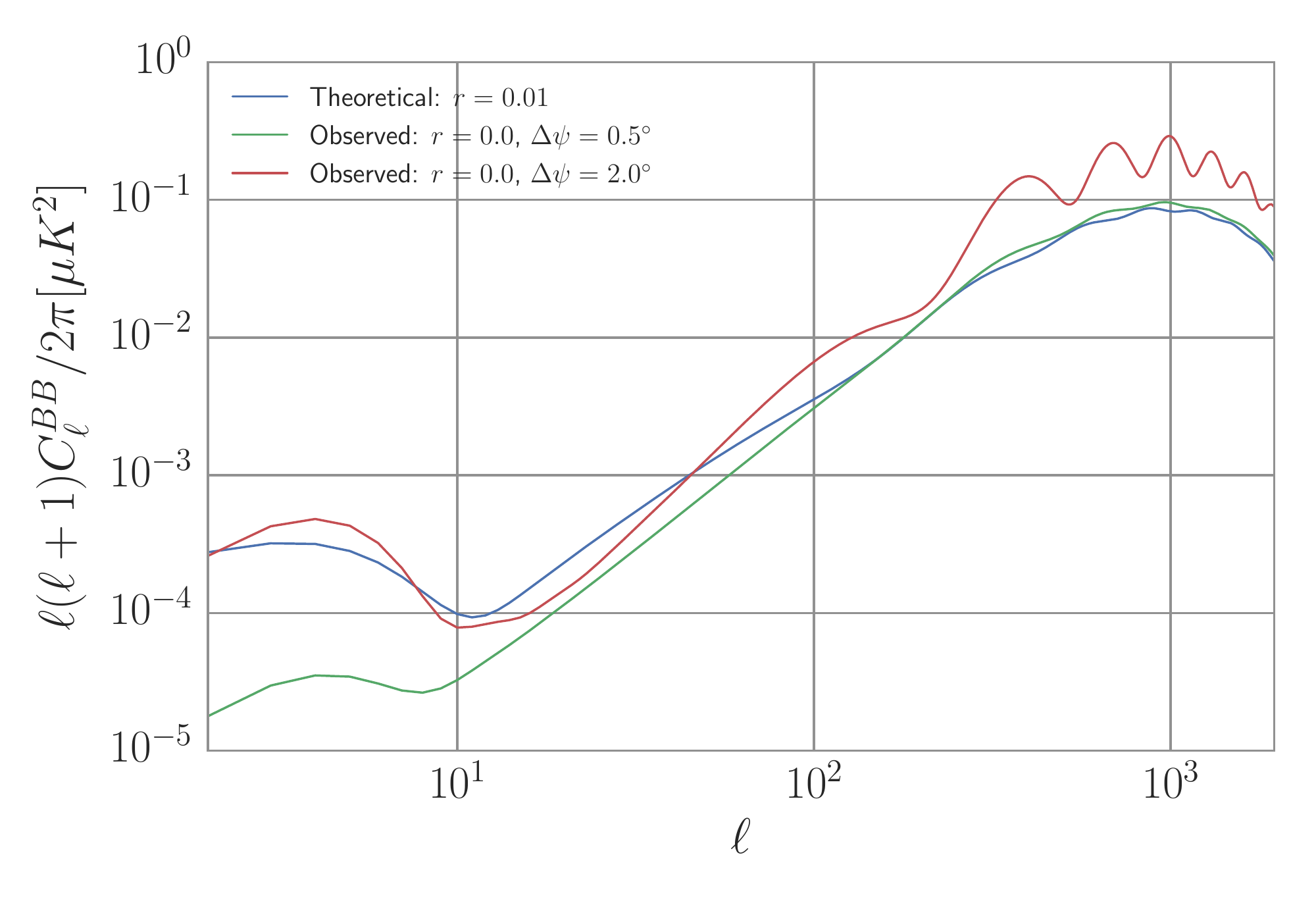}
    \caption{Rotated and theoretical $BB$ power spectra. The blue curve shows the theoretical CMB $BB$ power spectrum with $r=0.01$, including lensing. We compare this to the rotated $BB$ spectra, in the green and red curves, given a miscalibration of the telescope angle by $\Delta\psi=0.5^{\circ}$ and $2.0^{\circ}$, respectively. The rotated spectra were calculated using $r=0.0$, and thus consists of lensing and leaked $E-$ to $B$-modes only. For a misalignment of $\Delta\psi<0.5^{\circ}$, the $E$-mode leakage does not contribute significantly to the $BB$ spectrum until $\ell\gtrsim100$.}
    \label{fig:spurb}
\end{figure}

Current experiment systematic biases are generally larger than the potential foreground-induced bias. For example, BICEP2 measures a self-calibration angle of $\Delta\psi=-1.1^{\circ}$~\citep{bicep2III}, POLARBEAR measures $\Delta\psi=-1.08^{\circ}$ with a statistical uncertainty of $0.2^{\circ}$~\citep{polarbear_2014}, and ACTPOL constrains their polarisation offset angle to $-0.2\pm0.5^{\circ}$~\citep{naess_2014}. A $5\sigma$ detection of $r=0.01$ requires a polarisation angle uncertainty $<0.5^{\circ}$ for an otherwise ideal experiment with no other sources of systematic error. Accounting for other instrument systematics brings this requirement to $<0.2^{\circ}$~\citep{bock2006,odea,keatingself,johnson_2015}, which approaches the foreground bias for low-resolution experiments observing particularly dusty regions. Similarly, a miscalibration of the telescope angle by $\gtrsim0.5^{\circ}$ greatly biases the measurement of gravitational lensing of $E$-modes into $B$-modes~\citep{shimon2008}, as can be seen qualitatively in Fig.~\ref{fig:spurb}.
\begin{table}
\begin{center}
\begin{tabular}[!b]{ |c|c|c|c|c|c|c|c|c|c|c|c| } 
 \hline
  $\Delta\psi$ & $0.0^{\circ}$ & $0.2^{\circ}$ & $0.5^{\circ}$ & $1.0^{\circ}$ & $2.0^{\circ}$ \\ \hline
   $r$ & 0.000 & 0.0003 & 0.002 & 0.008 & 0.033 \\
 \hline
\label{tab:rstab}
\end{tabular}
\caption{Estimated spurious $r$ due to a rotation. We set $r$ as the ratio of the rotated $BB$ spectrum to the theoretical $r=1.0$ $BB$ spectrum at $\ell=80$.}
\end{center}
\end{table}

\section{Foreground Corrected Self-Calibration Method}
\label{secfore}
\begin{figure}
	\includegraphics[width=\columnwidth]{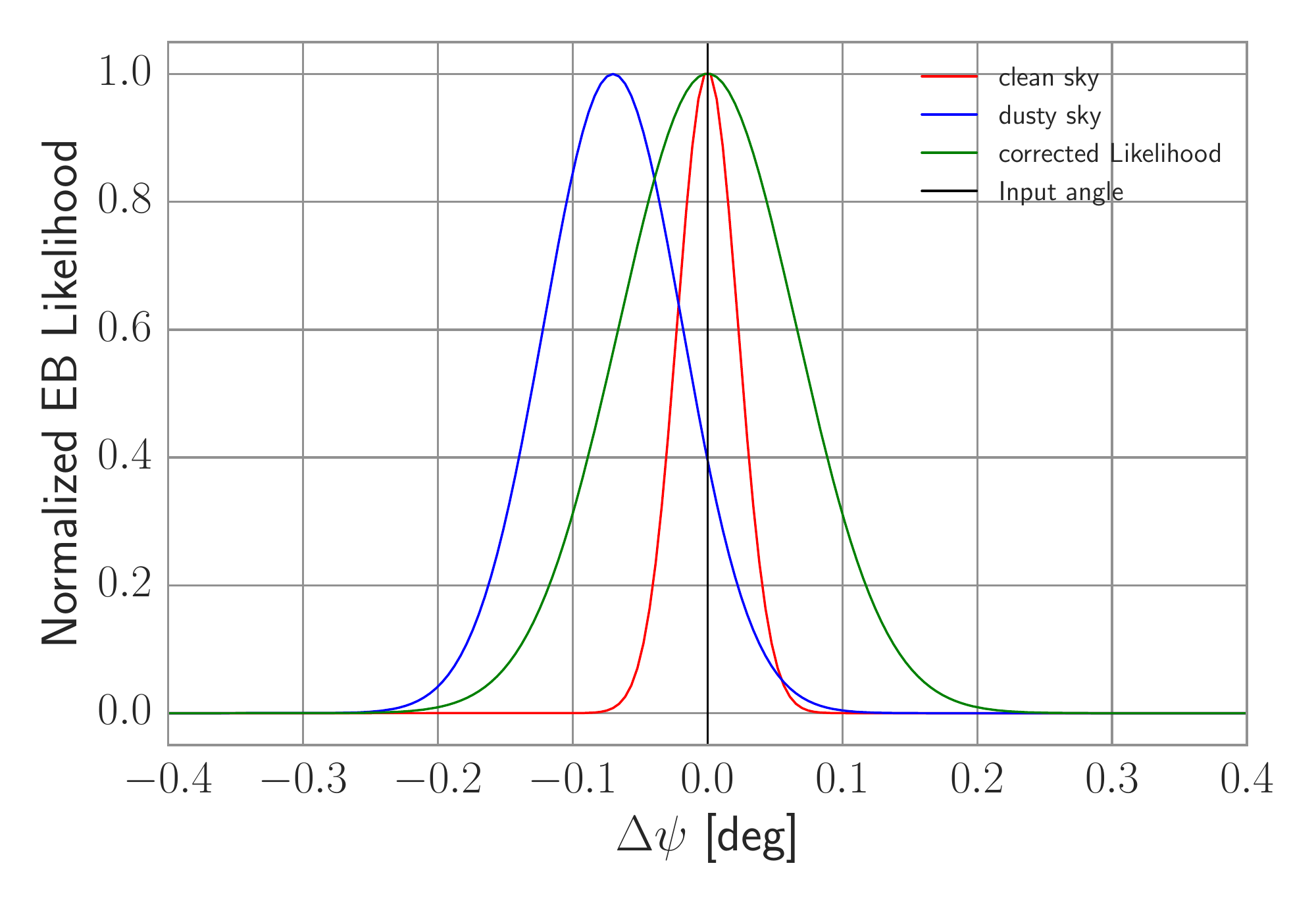}
    \caption{Likelihood corrected for foregrounds (see Section~\ref{secfore}). We use experiment configuration $E_{LR}$ and compare the uncorrected to the foreground-corrected likelihood. The red curve shows the likelihood without foregrounds for reference. The blue curve adds dust with $m=5$ into the rotated spectra using the original (uncorrected) likelihood (Equation~\ref{eq:likelihoods}). The green curve includes the same dust power but corrects for foregrounds in the likelihood (Equation~\ref{eq:dlikelihoods}). Including the dust in the likelihood eliminates the bias but increases the statistical uncertainty (see also Table~\ref{tab:full}).}
    \label{fig:EBfull}
\end{figure}

We incorporate foregrounds into the calibration method by including them explicitly in the likelihood functions as given by Equations~\ref{eq:dlikelihoods} and~\ref{eq:dlikelihoods2}. This has the effect of eliminating the bias but increasing the uncertainty on the calibration angle, as shown in Fig.~\ref{fig:EBfull}. We marginalize over foreground amplitudes assuming a fixed index foreground power law spectrum, although this can be straightforwardly generalized:
\begin{flalign}
\mathcal{L}_{EB}(\Delta\psi, \textit{\textbf{A}}^{\prime}) \propto & \exp -\Bigg[\sum_{\ell}\Bigg(\hat{C}^{EB}_{\ell} - \cos(4\Delta\psi) C^{\prime fg,EB}_{\ell} \nonumber \\ & + \frac{1}{2}\sin(4\Delta\psi)\Big(C^{EE}_{\ell} - C^{BB}_{\ell} \nonumber \\ & + C^{\prime fg,EE}_{\ell} - C^{\prime fg,BB}_{\ell}\Big)\Bigg)^2 \bigg/ 2\big(\delta \hat{C}^{EB}_{\ell} \big)^2\Bigg] \nonumber \\ & \times \exp -\Bigg[\frac{(A^{\prime EB}-A^{EB})^2}{2\sigma^2_{EB}} \nonumber \\ & +\frac{(A^{\prime EE}-A^{EE})^2}{2\sigma^2_{EE}}+\frac{(A^{\prime BB}-A^{BB})^2}{2\sigma^2_{BB}}\Bigg]
\label{eq:dlikelihoods}
\end{flalign}
\begin{flalign}
\mathcal{L}_{TB}(\Delta\psi, \textit{\textbf{A}}^{\prime}) \propto & \exp -\Bigg[\sum_{\ell} \Big(\hat{C}^{TB}_{\ell} - \cos(2\Delta\psi)C^{\prime fg,TB}_{\ell} \nonumber 
\\ & + \sin(2\Delta\psi)\big(C^{TE}_{\ell} + C^{\prime fg,TE}_{\ell}\big)\Big)^2\bigg/ 2\big(\delta\hat{C}^{TB}_{\ell}\big)^2\Bigg] \nonumber \\ & \times \exp -\Bigg[\frac{(A^{\prime TB}-A^{TB})^2}{2\sigma^2_{TB}} + \frac{(A^{\prime TE}-A^{TE})^2}{2\sigma^2_{TE}}\Bigg]
\label{eq:dlikelihoods2}
\end{flalign}
\begin{equation}
\mathcal{L}_{XB}(\Delta\psi) \propto \int d\textit{\textbf{A}}^{\prime} \mathcal{L}_{XB}(\Delta\psi, \textit{\textbf{A}}^{\prime}) \,.
\label{eq:marg}
\end{equation}
\noindent Here $C^{\prime fg,XY}_{\ell}$ represents the foreground power spectra determined by the amplitude $A^{\prime XY}$. We marginalize over the prime quantities in Equation~\ref{eq:marg}. The Gaussians are centred on the best-fitting foreground amplitude, $A^{XY}$, with variance $\sigma^2_{XY}$, as determined from 353~GHz (or other high frequency) data. Fig.~\ref{fig:EBfull} compares self-calibration results using the original and foreground-corrected likelihood functions. The corrected version accurately finds the calibration angle, with a slightly larger uncertainty due to the marginalization, as expected.   

\begin{table}
\begin{center}
\begin{tabular}{|c|c|c|c|} 
 \hline
  \multicolumn{2}{|c|}{Experiment Config.} & \multicolumn{2}{|c|}{$\Delta\psi$ [arcmin]} \\ \hline
  Dust Level $m$ & Corrected $\mathcal{L}$ & $EB$ & $TB$ \\ \hline
  $0$ & No & $0.0\pm1.4$ & $0.0\pm2.6$ \\
  $5$ & No & $-4.2\pm3.1$ & $-1.3\pm5.0$ \\
  $5$ & Yes & $0.0\pm3.9$ & $0.0\pm5.3$ \\
 \hline
\end{tabular}
\caption{Likelihood corrected for foregrounds (see Fig.~\ref{fig:EBfull}). Simulated misalignment $\Delta\psi_{\rm in}=0.0^{\circ}$ for experiment $E_{LR}$ and $m=0$ or $5$. Using the full likelihood calculation recovers the correct calibration angle as if there were no dust, but has larger uncertainty.}
\label{tab:full}
\end{center}
\end{table}

\section{Discussion}
\label{secdisc}
Unmitigated foreground interference can bias and appreciably reduce the utility of the CMB self-calibration method because a foreground biased polarisation angle will generate spurious $B$-mode power. We consider only dust in this paper, however, at lower frequencies other polarized sources such as synchrotron will likewise bias and reduce the effectiveness of the self-calibration procedure. To account for polarized foreground signals one can either include them in the self-calibration likelihood function or subtract them in the map domain. A map domain foreground cleaning may require an iterative method between self-calibration and component separation, especially if combining data from multiple instruments. 

We note that, in principle, experiments should simultaneously estimate both the cosmological parameter values and the polarisation angle because the cosmological parameters used as inputs to the theoretical CMB power spectra have non-zero uncertainty. Additionally, the likelihood functions for $EB$ and $TB$ should be maximized simultaneously, although the use of two separate estimators provides a consistency check. 

It is important to note that primordial magnetic fields and cosmic birefringence should produce faint non-zero $EB$ and $TB$ cross-spectra~\citep{planckmag, polarbear_birefringe}. Because the self-calibration method minimizes the $EB$ and $TB$ correlation, it is difficult to both search for these signals and self-calibrate. Nevertheless some experiments are investigating ways to make this observation~\citep{polarbear_birefringe}. 

We conclude that experiments using the self-calibration procedure should be aware of the potential bias of non-zero $EB$ and $TB$ power due to foregrounds. CMB experiments using foreground monitors at frequencies far above or below the foreground minimum need to account for foreground contamination in the self-calibration procedure. Self-calibration for experiments with access to high-$\ell$ multipoles is robust to foreground contamination, as the foreground power spectra generally falls off as a power law. Low-resolution or low-$\ell$ experiments observing small sky fractions are vulnerable to foreground-induced biases.

\section*{Acknowledgements}
This work was partially supported by a Junior Fellow award from the Simons Foundation to JCH. We thank Raphael Flauger, Tobias Marriage, Amber Miller, and David Spergel for helpful conversations. 

\bibliographystyle{mnras}
\bibliography{paper.bib}

\appendix{}
\section{Dust $EB$ and $TB$ Data}
\label{sec:A}
For completeness, we list the amplitudes used in the $EB$ and $TB$ power-law spectra in Table~\ref{tab:dustamps}. These can be compared to the $EB$ and $TB$ spectra in fig B.2 and B.3 of~\cite{planck2013foregrounds}. Briefly, Planck measures $EB$ and $TB$ power at 353~GHz in the range $0-10$ and $0-100$ $\mu K^2$, respectively, depending on the sky fraction analysed. Those upper limits correspond to approximately $0.017$ and $0.17$ $\mu K^2$ when scaled to 150~GHz using the grey-body frequency dependence of dust emission~\citep{planck2015foregrounds}. Comparing these to Table~\ref{tab:dustamps} (note we multiplied the Table by 1000 to ease readability), we see that all our spectra are within those bounds except the extreme case where $m=5$ and $f_c=0.5$.  

\begin{figure}
	\includegraphics[width=\columnwidth]{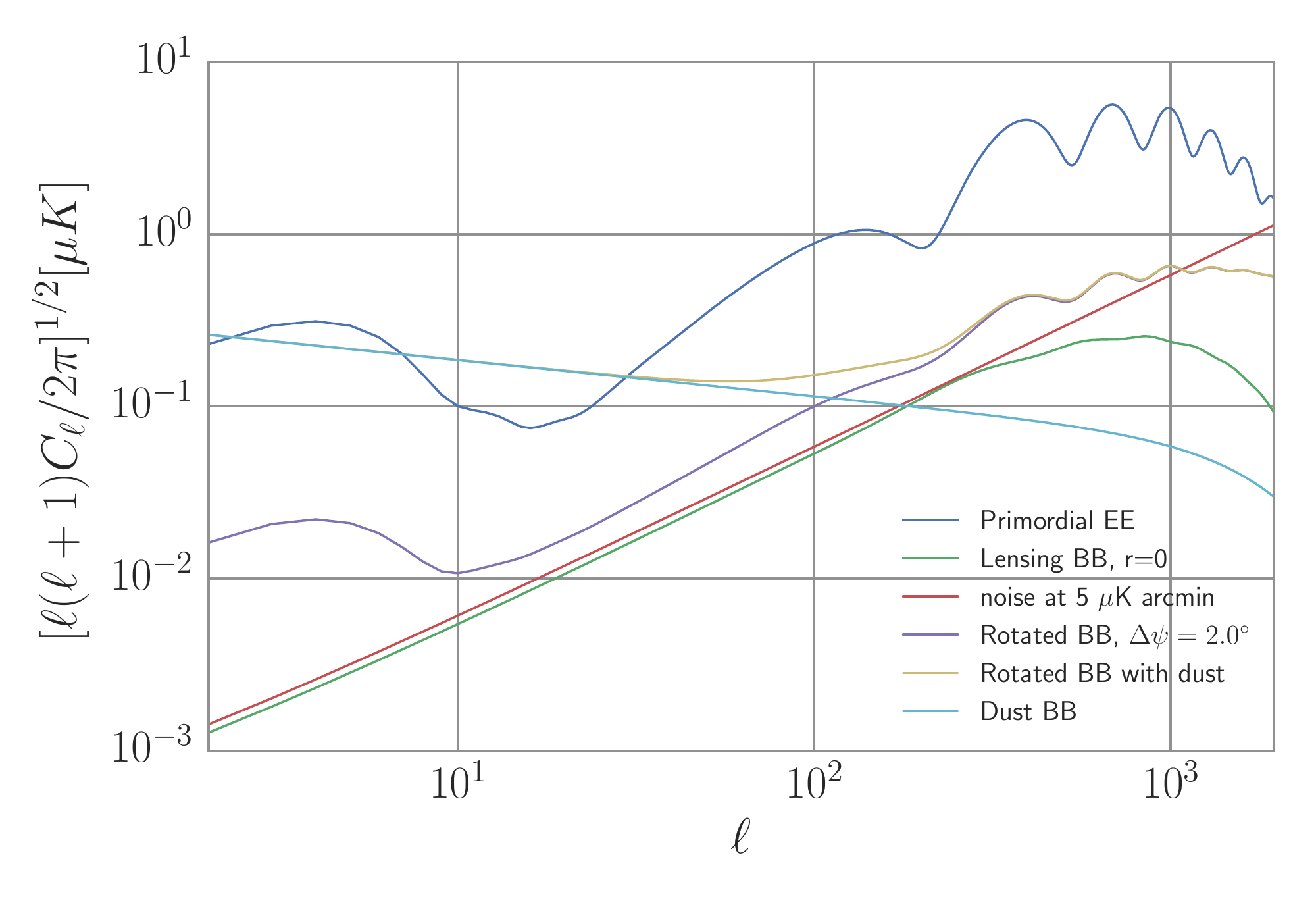}
    \caption{Theoretical and rotated amplitude spectra. CMB $EE$ and $BB$ amplitude spectra (square root of power spectra) with $r=0$ and rotated $BB$ spectra when including a telescope misalignment of $\Delta\psi_{\rm in}=2.0^{\circ}$. Also shown is the dust $BB$ spectrum and the noise amplitude spectrum given $\Delta_X=5 \mu K$ arcmin. The rotated spectra are shown both before and after adding dust to the sky in purple and gold lines, respectively. At low multipoles, $\ell\lesssim100$, the dust contributes significantly to the rotated $BB$ spectrum.}
    \label{fig:rotatedpower}
\end{figure}

Lastly, we reproduce fig. 2 of~\cite{keatingself} using our data sets and show the resulting rotated $BB$ spectra in Figure~\ref{fig:rotatedpower}. The rotated $BB$ spectra follows the dust spectra for $\ell\lesssim100$ and then follows the leaked $EE$ component for $\ell\gtrsim300$. 

\begin{table}
\begin{center}
\begin{tabular}{|c|c|c|} 
 \hline
  Dust Data Set & \multicolumn{2}{|c|}{Dust Power $[\ell(\ell+1)/2\pi$ $\mu K^2]\times1000$} \\ \hline
  Dust Params & $C^{dust,EB}_{\ell=80}$ & $C^{dust,TB}_{\ell=80}$ \\  \hline
  Measured & $0.39\pm3.5$ & $5.66\pm29$ \\
  Best-Fitting & $0.64\pm3.2$ & $-11.8\pm24$ \\ \hline
  $m=5$ & $3.2$ & $-59.2$ \\
  $m=10$ & $6.4$ & $-118$ \\ \hline
  $m=5$, $f_c=0.01$ & $0.86$ & $17.0$ \\
  $m=5$, $f_c=0.1$ & $8.6$ & $171$ \\
  $m=5$, $f_c=0.5$ & $43$ & $853$ \\
 \hline
\end{tabular}
\caption{Dust $EB$ and $TB$ Power (see Section~\ref{secdata}). We show the dust cross-spectra at $\ell=80$ (multiplied by 1000 for ease of reading) for the three data sets we use in this paper (see Section~\ref{secdata}). The best-fitting row refers to the amplitude of the power-law fit to all four band-powers (normalized at $\ell=80$), whereas the measured row refers to the band-power measured at $\ell=80$. The dust cross-spectra are currently not well-constrained and are consistent with zero in the BICEP2 region. We thus use these data sets to represent other possible measurements of the $EB$ and $TB$ dust cross-spectra, which are consistent with the bounds set by Planck, except for the $f_c=0.5$ case.}
\label{tab:dustamps}
\end{center}
\end{table}

\bsp	
\label{lastpage}
\end{document}